\documentclass[%
 reprint,
superscriptaddress,
 amsmath,amssymb,
 aps,
floatfix,
]{revtex4-2}

\usepackage{graphicx}% Include figure files
\usepackage{dcolumn}% Align table columns on decimal point
\usepackage{bm}% bold math
\usepackage{lineno}

\usepackage[version=3]{mhchem}
\usepackage{xspace}
\usepackage{multirow}
\newcommand{\E}{\ensuremath{\eta}~meson\xspace}
\newcommand{\Es}{\ensuremath{\eta}~mesons\xspace}
\newcommand{\Ox}{$\rm{{}^{16}O}$\xspace}

\begin{document}

\preprint{APS/123-QED}

\title{Search for proton decay via $p\rightarrow{e^+\eta}$ and $p\rightarrow{\mu^+\eta}$ with a 0.37 Mton-year exposure of Super-Kamiokande}

%\author{Natsumi Taniuchi}
%\affiliation{Your affiliation}

\newcommand{\AFFicrr}{\affiliation{Kamioka Observatory, Institute for Cosmic Ray Research, University of Tokyo, Kamioka, Gifu 506-1205, Japan}}
\newcommand{\AFFkashiwa}{\affiliation{Research Center for Cosmic Neutrinos, Institute for Cosmic Ray Research, University of Tokyo, Kashiwa, Chiba 277-8582, Japan}}
\newcommand{\AFFicrronly}{\affiliation{Institute for Cosmic Ray Research, University of Tokyo, Kashiwa, Chiba 277-8582, Japan}}
\newcommand{\AFFipmu}{\affiliation{Kavli Institute for the Physics and
Mathematics of the Universe (WPI), The University of Tokyo Institutes for Advanced Study,
University of Tokyo, Kashiwa, Chiba 277-8583, Japan }}
\newcommand{\AFFmad}{\affiliation{Department of Theoretical Physics, University Autonoma Madrid, 28049 Madrid, Spain}}
\newcommand{\AFFubc}{\affiliation{Department of Physics and Astronomy, University of British Columbia, Vancouver, BC, V6T1Z4, Canada}}
\newcommand{\AFFbu}{\affiliation{Department of Physics, Boston University, Boston, MA 02215, USA}}
\newcommand{\AFFuci}{\affiliation{Department of Physics and Astronomy, University of California, Irvine, Irvine, CA 92697-4575, USA }}
\newcommand{\AFFcsu}{\affiliation{Department of Physics, California State University, Dominguez Hills, Carson, CA 90747, USA}}
\newcommand{\AFFcnm}{\affiliation{Institute for Universe and Elementary Particles, Chonnam National University, Gwangju 61186, Korea}}
\newcommand{\AFFduke}{\affiliation{Department of Physics, Duke University, Durham NC 27708, USA}}
\newcommand{\AFFfukuoka}{\affiliation{Junior College, Fukuoka Institute of Technology, Fukuoka, Fukuoka 811-0295, Japan}}
\newcommand{\AFFgifu}{\affiliation{Department of Physics, Gifu University, Gifu, Gifu 501-1193, Japan}}
\newcommand{\AFFgist}{\affiliation{GIST College, Gwangju Institute of Science and Technology, Gwangju 500-712, Korea}}
\newcommand{\AFFuh}{\affiliation{Department of Physics and Astronomy, University of Hawaii, Honolulu, HI 96822, USA}}
\newcommand{\AFFicl}{\affiliation{Department of Physics, Imperial College London , London, SW7 2AZ, United Kingdom }}
\newcommand{\AFFkek}{\affiliation{High Energy Accelerator Research Organization (KEK), Tsukuba, Ibaraki 305-0801, Japan }}
\newcommand{\AFFkobe}{\affiliation{Department of Physics, Kobe University, Kobe, Hyogo 657-8501, Japan}}
\newcommand{\AFFkyoto}{\affiliation{Department of Physics, Kyoto University, Kyoto, Kyoto 606-8502, Japan}}
\newcommand{\AFFliv}{\affiliation{Department of Physics, University of Liverpool, Liverpool, L69 7ZE, United Kingdom}}
\newcommand{\AFFmiyagi}{\affiliation{Department of Physics, Miyagi University of Education, Sendai, Miyagi 980-0845, Japan}}
\newcommand{\AFFnagoya}{\affiliation{Institute for Space-Earth Environmental Research, Nagoya University, Nagoya, Aichi 464-8602, Japan}}
\newcommand{\AFFkmi}{\affiliation{Kobayashi-Maskawa Institute for the Origin of Particles and the Universe, Nagoya University, Nagoya, Aichi 464-8602, Japan}}
\newcommand{\AFFpol}{\affiliation{National Centre For Nuclear Research, 02-093 Warsaw, Poland}}
\newcommand{\AFFsuny}{\affiliation{Department of Physics and Astronomy, State University of New York at Stony Brook, NY 11794-3800, USA}}
\newcommand{\AFFokayama}{\affiliation{Department of Physics, Okayama University, Okayama, Okayama 700-8530, Japan }}
\newcommand{\AFFosaka}{\affiliation{Department of Physics, Osaka University, Toyonaka, Osaka 560-0043, Japan}}
\newcommand{\AFFox}{\affiliation{Department of Physics, Oxford University, Oxford, OX1 3PU, United Kingdom}}
\newcommand{\AFFqmul}{\affiliation{School of Physics and Astronomy, Queen Mary University of London, London, E1 4NS, United Kingdom}}
\newcommand{\AFFregina}{\affiliation{Department of Physics, University of Regina, 3737 Wascana Parkway, Regina, SK, S4SOA2, Canada}}
\newcommand{\AFFseoul}{\affiliation{Department of Physics, Seoul National University, Seoul 151-742, Korea}}
\newcommand{\AFFsheff}{\affiliation{Department of Physics and Astronomy, University of Sheffield, S3 7RH, Sheffield, United Kingdom}}
\newcommand{\AFFshizuokasc}{\affiliation{Department of Informatics in
Social Welfare, Shizuoka University of Welfare, Yaizu, Shizuoka, 425-8611, Japan}}
\newcommand{\AFFstfc}{\affiliation{STFC, Rutherford Appleton Laboratory, Harwell Oxford, and Daresbury Laboratory, Warrington, OX11 0QX, United Kingdom}}
\newcommand{\AFFskk}{\affiliation{Department of Physics, Sungkyunkwan University, Suwon 440-746, Korea}}
\newcommand{\AFFtokyo}{\affiliation{The University of Tokyo, Bunkyo, Tokyo 113-0033, Japan }}
\newcommand{\AFFtodai}{\affiliation{Department of Physics, University of Tokyo, Bunkyo, Tokyo 113-0033, Japan }}
\newcommand{\AFFtit}{\affiliation{Department of Physics,Tokyo Institute of Technology, Meguro, Tokyo 152-8551, Japan }}
\newcommand{\AFFtus}{\affiliation{Department of Physics, Faculty of Science and Technology, Tokyo University of Science, Noda, Chiba 278-8510, Japan }}
\newcommand{\AFFtoronto}{\affiliation{Department of Physics, University of Toronto, ON, M5S 1A7, Canada }}
\newcommand{\AFFtriumf}{\affiliation{TRIUMF, 4004 Wesbrook Mall, Vancouver, BC, V6T2A3, Canada }}
\newcommand{\AFFtokai}{\affiliation{Department of Physics, Tokai University, Hiratsuka, Kanagawa 259-1292, Japan}}
\newcommand{\AFFtsinghua}{\affiliation{Department of Engineering Physics, Tsinghua University, Beijing, 100084, China}}
\newcommand{\AFFynu}{\affiliation{Department of Physics, Yokohama National University, Yokohama, Kanagawa, 240-8501, Japan}}
\newcommand{\AFFllr}{\affiliation{Ecole Polytechnique, IN2P3-CNRS, Laboratoire Leprince-Ringuet, F-91120 Palaiseau, France }}
\newcommand{\AFFbari}{\affiliation{ Dipartimento Interuniversitario di Fisica, INFN Sezione di Bari and Universit\`a e Politecnico di Bari, I-70125, Bari, Italy}}
\newcommand{\AFFnapoli}{\affiliation{Dipartimento di Fisica, INFN Sezione di Napoli and Universit\`a di Napoli, I-80126, Napoli, Italy}}
\newcommand{\AFFroma}{\affiliation{INFN Sezione di Roma and Universit\`a di Roma ``La Sapienza'', I-00185, Roma, Italy}}
\newcommand{\AFFpadova}{\affiliation{Dipartimento di Fisica, INFN Sezione di Padova and Universit\`a di Padova, I-35131, Padova, Italy}}
\newcommand{\AFFkeio}{\affiliation{Department of Physics, Keio University, Yokohama, Kanagawa, 223-8522, Japan}}
\newcommand{\AFFwinnipeg}{\affiliation{Department of Physics, University of Winnipeg, MB R3J 3L8, Canada }}
\newcommand{\AFFkcl}{\affiliation{Department of Physics, King's College London, London, WC2R 2LS, UK }}
\newcommand{\AFFwarwick}{\affiliation{Department of Physics, University of Warwick, Coventry, CV4 7AL, UK }}
\newcommand{\AFFral}{\affiliation{Rutherford Appleton Laboratory, Harwell, Oxford, OX11 0QX, UK }}
\newcommand{\AFFwu}{\affiliation{Faculty of Physics, University of Warsaw, Warsaw, 02-093, Poland }}
\newcommand{\AFFbcit}{\affiliation{Department of Physics, British Columbia Institute of Technology, Burnaby, BC, V5G 3H2, Canada }}
\newcommand{\AFFtohoku}{\affiliation{Department of Physics, Faculty of Science, Tohoku University, Sendai, Miyagi, 980-8578, Japan }}
\newcommand{\AFFicise}{\affiliation{Institute For Interdisciplinary Research in Science and Education, ICISE, Quy Nhon, 55121, Vietnam }}
\newcommand{\AFFilance}{\affiliation{International Laboratory for Astrophysics, Neutrino and Cosmology Experiment, Kashiwa, Chiba 277-8582, Japan}}
\newcommand{\AFFibs}{\affiliation{Center for Underground Physics, Institute for Basic Science (IBS), Daejeon, 34126, Korea}}
\newcommand{\AFFglasgow}{\affiliation{School of Physics and Astronomy, University of Glasgow, Glasgow, Scotland, G12 8QQ, United Kingdom}}
\newcommand{\AFFoecu}{\affiliation{Media Communication Center, Osaka Electro-Communication University, Neyagawa, Osaka, 572-8530, Japan}}
\newcommand{\AFFminn}{\affiliation{School of Physics and Astronomy, University of Minnesota, Minneapolis, MN  55455, USA}}
\newcommand{\AFFsilesia}{\affiliation{August Che\l{}kowski Institute of Physics, University of Silesia in Katowice, 75 Pu\l{}ku Piechoty 1, 41-500 Chorz\'{o}w, Poland}}
\newcommand{\AFFtoyama}{\affiliation{Faculty of Science, University of Toyama, Toyama City, Toyama 930-8555, Japan}}
\newcommand{\AFFbmcc}{\affiliation{Science Department, Borough of Manhattan Community College / City University of New York, New York, New York, 1007, USA.}}

\AFFicrr
\AFFkashiwa
\AFFicrronly
\AFFmad
\AFFbu
\AFFbcit
\AFFuci
\AFFcsu
\AFFcnm
\AFFduke
\AFFllr
\AFFgifu
\AFFgist
\AFFglasgow
\AFFuh
\AFFibs
\AFFicise
\AFFicl
\AFFbari
\AFFnapoli
\AFFpadova
\AFFroma
\AFFilance
\AFFkeio
\AFFkek
\AFFkcl
\AFFkobe
\AFFkyoto
\AFFliv
\AFFminn
\AFFmiyagi
\AFFnagoya
\AFFkmi
\AFFpol
\AFFsuny
\AFFokayama
\AFFoecu
\AFFox
\AFFral
\AFFseoul
\AFFsheff
\AFFshizuokasc
\AFFsilesia
\AFFstfc
\AFFskk
\AFFtohoku
\AFFtokai
\AFFtokyo
\AFFtodai
\AFFipmu
\AFFtit
\AFFtus
\AFFtoyama
\AFFtoronto
\AFFtriumf
\AFFtsinghua
\AFFwu
\AFFwarwick
\AFFwinnipeg
\AFFynu

\author{N.~Taniuchi}
\altaffiliation{Present Address: Department of Physics, Cavendish Laboratory, University of Cambridge, Cambridge, CB3 0HE, United Kingdom}
%\thanks{Department of Physics, Cavendish Laboratory, University of Cambridge, Cambridge, CB3 0HE, United Kingdom}
\AFFtodai
%\AFFcam

%%%%%%%%%%%%%%%%%%%%%%%%%%%%%%%%%%%%%%%%%%%%%%%%%%%%%%%%%%%%%%%%%%%%
%ICRR
\author{K.~Abe}
\AFFicrr
\AFFipmu
\author{S.~Abe}
\author{Y.~Asaoka}
\author{C.~Bronner}
\author{M.~Harada}
\AFFicrr
\author{Y.~Hayato}
\AFFicrr
\AFFipmu
\author{K.~Hiraide}
\AFFicrr
\AFFipmu
\author{K.~Hosokawa}
\AFFicrr
\author{K.~Ieki}
\author{M.~Ikeda}
\AFFicrr
\AFFipmu
% ...

%%%%%%%%%%%%%%%%%%%%%%%%%%%%%%%%%%%%%%%%%%%%%%%%%%%%%%%%%%%%%%%%%%%%%

\collaboration{The Super-Kamiokande Collaboration}
\noaffiliation

%Nucleon is spelled out.

\date{\today}% It is always \today, today,

\begin{abstract}
%% Grand Unified Theories explain the unification of the electromagnetic, weak, and strong forces and most of them predict protons to decay into lighter particles.
A search for proton decay into $e^+/\mu^+$ and a $\eta$ meson has been performed using data from a 0.373 Mton$\cdot$year exposure (6050.3 live days) of Super-Kamiokande.
Compared to previous searches this work introduces an improved model of the intranuclear $\eta$ interaction cross section, resulting in a factor of two reduction in uncertainties from this source and $\sim$10\% increase in signal efficiency. 
%%The cross sections of $\eta$ nuclear effect are improved compared to previous work, resulting in reducing their uncertainties around by a factor of two. 
%%We analyze the 
No significant data excess was found above the expected number of atmospheric neutrino background events resulting in no indication of proton decay into either mode.
Lower limits on the proton partial lifetime of $1.4\times\mathrm{10^{34}~years}$ for $p\rightarrow e^+\eta$ and $7.3\times\mathrm{10^{33}~years}$ for $p\rightarrow \mu^+\eta$ at the 90$\%$ C.L. were set.
These limits are around 1.5 times longer than our previous study and are the most stringent to date. 
\end{abstract}

\maketitle

\section{\label{sec:intro}Introduction}

%The Standard Model of particle physics has succeeded in explaining most of the experimental results with accuracy, however it also appears to be an incomplete description.
%It fails to provide a sufficient explanation for the quantization of electric charge or the convergence of the electromagnetic, weak, and strong interaction couplings at high energy scales.
%Grand Unified Theories (GUTs) have been proposed to address those issues.
%The main idea of GUTs is that the gauge symmetries of $\mathrm{SU(3)}_c\times\mathrm{SU(2)}_L\times\mathrm{U(1)}_Y$ of the Standard Model are incorporated into a larger single symmetry, giving a variety to the theories based on their gauge groups such as SU(5) \cite{Georgi1974} or SO(10) \cite{Fritzschandpeterminkowski1975}.

While the Standard Model of particle physics successfully explains most experimental data with exceptional precision it is not a complete description of nature.
Notably it does not offer sufficient explanations for the observed quantization of electric charge nor for the convergence of the electromagnetic, weak, and strong interaction couplings at high energy scales.
To address those questions, Grand Unified Theories (GUTs) extend the Standard Model by incorporating its 
$\mathrm{SU(3)}_c\times\mathrm{SU(2)}_L\times\mathrm{U(1)}_Y$ 
gauge symmetry into larger symmetry groups, such as SU(5) \cite{Georgi1974} or SO(10)~\cite{Fritzschandpeterminkowski1975}.
Most GUTs permit new interactions where leptons and quarks transform into each other and thereby induce baryon-number-violating proton decays with considerably long lifetimes.
Though the gauge couplings in GUTs are estimated to unify near $\sim10^{16}~\textrm{GeV}$, an energy too high to be reached by accelerators, the observation of proton decay would be an experimental verification of such high energy theories.

%a large water Cherenkov detector. 

While the predicted lifetime and products of proton decay depend on the GUT model, many are expected to be observable with the Super-Kamiokande (SK) experiment.
The $p\rightarrow l^+\pi^0$ $(l=e^+, \mu^+)$ and $p\rightarrow \bar{\nu}K^+$ modes, for example, are favored decays in many nonsupersymmetric and TeV-scale supersymmetric GUTs \cite{Sakai1982, Weinberg1982}, both of which have been tested at SK.
No significant signal has been observed in either~\cite{Takenaka2020, Abe2014a} and the world's most stringent limits, corresponding to lifetimes of the order of $10^{34}$ years, have been set.
The $p\rightarrow l^+\pi^0$ modes are readily probed at SK due to its large size and highly efficient detection of the decay products; two $\gamma$-rays from the $\pi^0$ decay and a single $l^+$, which is predicted to be above Cherenkov threshold, can be observed clearly.
However, some models~\cite{Machacek1979, Gavela1981, Donoghue1980, Buccella1989} suggest that the decay rates of channels with heavy non-strange mesons ($\eta$, $\rho$, $\omega$) are comparable in magnitude to those with pions.
In particular, the $p\rightarrow l^+\eta$ decay generates a similar signal to that from $p\rightarrow l^+\pi^0$, where the $\eta$ meson mainly decays into two $\gamma$-rays, resulting in a relatively high detection efficiency.
%Proton decays are supposed to occur with such an extremely low probability that the number of expected events is one or less for a single channel within the full SK dataset of more than 20 years.
%Therefore, it is crucial to survey comprehensively across multiple decay modes to enhance the chance of detection.
It is essential to search for such possibilities as well, since it is not known which proton decay mode is dominant and since the observation of multiple modes can be used to determine the GUT gauge group.

%%In scenarios where the anticipated occurrence in any channel with an extremely long lifetime is one event or less with the full SK dataset, it is crucial to explore various decay modes to increase the chance of detection.

Our most recent search for the $p\rightarrow l^+\eta$ decay used  0.316 Mton$\cdot$years of SK data \cite{Abe2017} and 
concluded with no positive
observations, setting lower limits on the partial lifetime of $\tau/B(p\rightarrow e^+\eta) > 1.0\times 10^{34}~\mathrm{years}$ and $\tau/B(p\rightarrow \mu^+\eta) > 4.7\times 10^{33}~\mathrm{years}$ at 90$\%$ confidence level ($\textrm{C.L.})$.
This paper describes improved searches for these two modes using a total  exposure of 0.373 Mton$\cdot$years, corresponding to a $\sim$18$\%$ increase in exposure from the SK-IV period.
The present work additionally includes an improved estimation of the intranuclear $\eta$ interaction cross sections and their accompanying uncertainties.

\section{\label{sec:SK}Super-Kamiokande Detector}

Super-Kamiokande is a large water Cherenkov detector located 1,000~m beneath Mt.~Ikenoyama (2700 m.w.e.), in the Kamioka-mine in Hida-city, Gifu Prefecture, Japan.
The detector is a cylindrical stainless-steel tank with a 39.3~m diameter and a 41.4~m height that is filled with 50~ktons of ultrapure water.
It is separated into two sections: the inner detector (ID) and the outer detector (OD).
The ID forms the main target mass and has a diameter of 33.8~m and a height of 36.2~m. It is viewed by more than 11,000 20-inch photomultiplier tubes (PMTs), which face  inwards. 
On the other hand, the primary functions of the OD are to veto
cosmic ray muons and to serve as a shield against radiation coming from the rock around the detector.
The OD is a 2~m-thick cylindrical annulus veiewed by 1,885 outward-facing 8-inch PMTs. 
The detector is described in more detail in \cite{Fukuda2003,ABE2014253}.

%%corresponding to around $40\%$ coverage.
%The OD has been mainly used for rejecting cosmic muon events as a veto detector and also as a shield against $\gamma$-rays from the surrounding rocks.

Up until 2018, the SK data are divided into four periods: SK-I, II, III, and IV. 
The SK-I period started on April 1st 1996 and ran with a photocathode coverage of 40$\%$ until July 2001.
Due to a chain reaction implosion in November 2001 the SK-II period was operated with 5,182 ID PMTs, corresponding to 19$\%$ photocathode coverage, from October 2002 to October 2005.
After replenishing the missing PMTs, data taking started again at the nominal 40$\%$ coverage from June 2006 to September 2008 (SK-III).
In September 2008, new front-end electronics \cite{Nishino2009a} were installed to start the SK-IV period, which ended in May 2018.
This upgrade allows events to be recorded with no dead time, thereby enabling the detection of $\gamma$-rays from neutron capture on a hydrogen nucleus.
 %% This technique contributed to more dedicated reduction of atmospheric neutrino backgrounds.
In this study, all available data from SK-I to SK-IV are utilized.
A total 0.373 Mton$\cdot$year exposure from 91.5, 49.1, 31.8, and 199.4 kton$\cdot$years of the SK-I, II, III, and IV periods, respectively, is used.

\section{\label{sec:sim}Simulation}

\subsection{\label{sec:sim_pdk}Proton Decay MC}

Dedicated proton decay Monte Carlo (MC) samples for $p\rightarrow l^+\eta$ modes are generated for the estimation of signal detection efficiencies.
Inside SK, proton decay events are assumed to occur both in hydrogen nuclei as free protons and in oxygen nuclei as bound protons in a ratio of one to four in the $\rm{\ce{H_2O}}$ molecule.
%The probabilities of proton decays are considered as the same regardless of their initial states.
Protons are considered to decay with the same probability regardless of their initial states.
The $p\rightarrow l^+\eta$ proton decay events are treated as two body decays. 
While hydrogen nuclei (free protons) are stationary and do not interact with other nucleons, bound protons in oxygen nuclei are subject to the effects of Fermi motion, nuclear binding energy, and correlated momentum effects with surrounding nucleons.
Furthermore, $\eta$ mesons emitted by proton decays can interact with nucleons, being absorbed or scattered, prior to exiting the nucleus.
The decaying proton's position in the nucleus is given by the Woods-Saxon nuclear density model~\cite{Woods1954}.

%the $\eta$ mesons? OR $\eta$s?

The Fermi motion and nuclear binding energy in \Ox~are simulated based on electron scattering experiments on ${}^{12}\rm{C}$~\cite{Nakamura1976}.
%%and calculated nucleon momentum and nuclear binding energy in ${}^{12}\rm{C}$ nucleus 
%%from an
The \Ox nuclear binding energies are $39.0~\rm{MeV}$ for the $s$-state and $15.5~\rm{MeV}$ for the $p$-state protons.
In the simulation, an effective mass of the proton is introduced by subtracting the binding energy from the proton rest mass. 
Decaying protons can be affected by the other nucleons in the same \Ox nucleus due to wavefunction overlap (correlated decay).
This effect is predicted to occur in $\sim 10\%$  of protons~\cite{Yamazaki1999}, resulting in an effective three-body decay with a recoiling nucleon, which carries away momentum and produces a tail in the lower mass region of the proton mass distribution.
%The decaying protons can be affected by the other nucleons in the same \Ox nucleus by their wave function overlapping with a predicted probability of $\sim 10\%$ \cite{Yamazaki1999}.
%In this case, the proton decays as a three-body decay with a recoiling nucleon.
Finally, the \Es~generated by bound protons in the \Ox~nucleus can interact with nucleons while escaping from the nucleus ($\eta$ nuclear effect, presented in detail in Section~\ref{sec:sim_nuc}). 
Each of these effects alter final kinematics of the decay particles relative to those from free proton decays.

%% Details of the $\eta$ nuclear effect are presented in Section.

%The decaying proton position in \Ox~nucleus is given by the Woods-Saxon nuclear density model \cite{Woods1954} as
%\begin{eqnarray}
%    \rho(r) = \frac{\rho(0)}{1+\exp{(\frac{r-a}{b})}}\label{density}~,
%\end{eqnarray}
%where $\rho(0)$ is the average density of nuclei, $a$ and $b$ are the maximum radius and the surface thickness of \Ox~nucleus, respectively.
%The \Es~are considered to be emitted from this position. The details of the $\eta$ nuclear effect are discussed in Section~\ref{sec:sim_nuc}.

\subsection{\label{sec:sim_nuc}$\eta$ Nuclear Effect}

$\eta$ mesons emitted from \Ox proton decays can interact with nucleons while travelling through the nucleus and form several baryon resonance states in the process.
Of these, the $S_{11}(1535)$ resonance state has been studied exclusively due to the property that it exists slightly above the $\eta-N$ production threshold, thereby providing a large branching ratio for $S_{11}(1535)\rightarrow N+\eta$ \cite{Krusche2015, Robig-Landau1996}.
Therefore, in this study the $\eta$ nuclear effect is evaluated through $S_{11}(1535)$ resonance state.

Since this nuclear effect plays a crucial role in proton decay searches due to its impact on the number of observable \Es, it is important to derive a reliable estimation of the cross section for this reaction. 
In our previous study \cite{Nishino2012}, the cross section of the $\eta$ nuclear effect, $\sigma_\mathrm{nuc}$, was calculated with the Breit-Wigner formula \cite{Breit1936}.
Experimental \E photoproduction data~\cite{Robig-Landau1996} was then used to assess its uncertainty:
The measured differential cross section for the $\eta$ photoproduction reaction $d\sigma_{\eta \mathrm{photo}}/dp$ \cite{Robig-Landau1996} was compared with the simulated value from the NEUT interaction generator \cite{Hayato2002, Hayato2009}, which includes the calculated $\eta$ nuclear effect cross section.  
%% $\sigma_\mathrm{nuc}$.
%from $\sigma_\mathrm{nuc}$.
%The process of \Es being absorbed within the oxygen nuclei brought the largest uncertainty in the previous search.

%This study, on the other hand, additionally makes use of an $\eta$ absorption cross section measurement, $\sigma_\mathrm{abs}$, which was directly extracted from $\eta$ photoproduction data.
This study, on the other hand, additionally makes use of $\eta$ absorption cross section measurement, $\sigma_\mathrm{abs}$, which was independently deduced from total $\eta$ photoproduction data on various targets \cite{Robig-Landau1996}.
%Here  $\sigma_\mathrm{nuc}$ and its uncertainty are estimated by adopting the least $\chi^2$ method to fit functional forms to the measured $\sigma_\mathrm{abs}$ from \cite{Robig-Landau1996}.
Here $\sigma_\mathrm{nuc}$ and its uncertainty are estimated by adopting the least $\chi^2$ method to fit functional forms to the measured $d\sigma_{\eta \mathrm{photo}}/dp$ and $\sigma_\mathrm{abs}$ from \cite{Robig-Landau1996}.
We assume that $\sigma_\mathrm{abs}\sim\sigma_\mathrm{nuc}$, with the 
difference between them coming from the inclusion of elastic scattering and multi-step interaction effects.
Both are expected to be minor, inducing only to a few percent difference as discussed in \cite{Robig-Landau1996}.

%This study, on the other hand, additionally makes use of $\eta$ absorption cross section measurement, $\sigma_\mathrm{abs}$, which was extracted by $d\sigma_{\eta \mathrm{photo}}/dp$ \cite{Robig-Landau1996} for the improved fitting of $\sigma_\mathrm{nuc}$.
%Here, $\sigma_\mathrm{abs}\sim\sigma_\mathrm{nuc}$ is assumed as the difference is the inclusion of effects on elastic scattering and multistep interactions, which can be assumed to play a minor role of up to a few percentage as discussed in \cite{Robig-Landau1996}.
%$\sigma_\mathrm{nuc}$ and its uncertainty are estimated by adopting the least $\chi^2$ method for both measured $d\sigma_{\eta \mathrm{photo}}/dp$ and $\sigma_\mathrm{abs}$ from \cite{Robig-Landau1996}.

\begin{figure}[b]
\includegraphics[width=\linewidth]{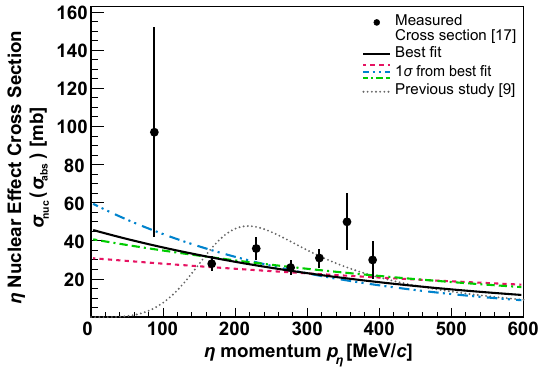}% Here is how to import EPS art
\caption{\label{fig:xs_nuc} Measured $\eta$ absorption cross section, $\sigma_\mathrm{abs}$, and fitted $\eta$ nuclear effect cross section, $\sigma_\mathrm{nuc}$. Experimental data for $\sigma_\mathrm{abs}$ from the Mainz experiment \cite{Robig-Landau1996} are shown as the black circles with statistical errors only. The black solid line shows the fitted $\sigma_\mathrm{nuc}$ at the best fit parameters $(a~[\textrm{mb}], b~[c\textrm{/MeV}]) = (46, 0.0023)$, while the red, green, and blue lines stand for that with representative parameters allowed at $\pm1\sigma$, $(a~[\textrm{mb}], b~[c\textrm{/MeV}]) = (31, 0.0010), (41, 0.0016),$ and $(60, 0.0032)$, respectively. The dashed line shows $\sigma_\mathrm{nuc}$ from the previous analysis \cite{Abe2017}, which was calculated according to the Breit-Wigner formula.}
\end{figure}

\begin{figure}[b]
\includegraphics[width=\linewidth]{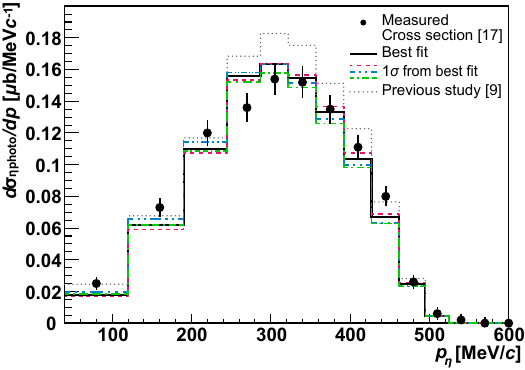}% Here is how to import EPS art
\caption{\label{fig:xs_ephoto} Measured and simulated differential $\eta$ photoproduction cross sections $d\sigma_{\eta \mathrm{photo}}/dp$.  Data from the Mainz experiment \cite{Robig-Landau1996} are shown as the black circles with statistical errors only.  The black solid line shows the $d\sigma_{\eta \mathrm{photo}}/dp$ calculated using the best fit parameters $(a~[\textrm{mb}], b~[c\textrm{/MeV}]) = (46, 0.0023)$ for $\sigma_\mathrm{nuc}$, while the red, green, and blue lines stand for that with representative parameters allowed at $\pm1\sigma$, $(a~[\textrm{mb}], b~[c\textrm{/MeV}]) = (31, 0.0010), (41, 0.0016),$ and $(60, 0.0032)$, respectively. The dashed line shows $\sigma_\mathrm{nuc}$ from the previous analysis \cite{Abe2017}.}
\end{figure}

%The fit for $\sigma_\mathrm{nuc}$ is performed with two different functions: $\sigma_\mathrm{nuc}=a$ and  $\sigma_\mathrm{nuc}=a e^{-b p_{\eta}}$, where $a$ and $b(b\neq0)$ are fit parameters and $p_\eta$ is the momentum of the \E in the laboratory frame.
We adopt an exponential parameterisation for $\sigma_\mathrm{nuc}$ given by 
$\sigma_\mathrm{nuc}=a e^{-b p_{\eta}}$, where $a$ and $b~(\geq0)$ are to be determined by the fit. 
Here $p_\eta$ is the momentum of the \E in the laboratory frame.
This form is then propagated through NEUT (version 5.4.0) to calculate $d\sigma_{\eta \mathrm{photo}}/dp$ as a function of $a$ and $b$. 
These parameters are varied during a simultaneous fit to both the $\eta$ absorbtion  (Figure~\ref{fig:xs_nuc}) and photoproduction cross section data (Figure~\ref{fig:xs_ephoto}) to determine the most compatibile cross section form.

Figure~\ref{fig:xs_nuc} shows the measured $\sigma_\mathrm{nuc}~(\sigma_\mathrm{abs})$ as well as our estimations at the best fit $(a~[\textrm{mb}], b~[c\textrm{/MeV}])=(46, 0.0023)$ and for parameters allowed at $1\sigma$.
Similarly, Figure~\ref{fig:xs_ephoto} shows the result for the $d\sigma_{\eta \mathrm{photo}}/dp$ cross section.
The additional 6\% systematic errors on the measured data points are not shown in the figures.
Compared to our previous parameterisation, shown as the dotten line in both figures, generally the data are better described by the new parameterisation. 
This improved $\sigma_\mathrm{nuc}$ estimation reduces the systematic error by up to 65$\%$ compared to the previous analysis as is described in Section~\ref{sec:result_syserr}.

%% The data are generally described better with the result of the new fit, with the exception of the bin at 250~MeV.

%must avoid using brackets? $\sigma_\mathrm{nuc}=a\exp{(-b\cdot p_{\eta})}$

\subsection{\label{sec:sim_atmnu}Atmospheric $\nu$ MC}

Backgrounds in the analysis are estimated using simulated atmospheric neutrino events corresponding to a 500-year exposure of the detector for each SK period.
The atmospheric neutrino flux used in the simulation is taken from the Honda calculation~\cite{Honda2007, Honda2011} and interactions are generated using the cross section models in NEUT~\cite{Hayato2002}, with updates described in~\cite{Abe2018}.
Events with neutral pions, which are the dominant background in the present analysis, are subject to interactions with nucleons in the \Ox~nucleus in a manner  similar to the $\eta$ nuclear effect. 
Pions can be scattered, absorbed, or charge exchanged ($\pi$ nuclear effect).
The cross sections for each of the processes are calculated by the NEUT cascade model~\cite{Hayato2002, Salcedo1988} with data from various $\pi-(p,n)$ and $\pi-N$ scattering experiments \cite{DePerio2011}.
Uncertainties on these processes are important error sources in the present measurement. 
Meson production from atmospheric neutrino interactions in water, including $\pi$ and $\eta$, are simulated in NEUT by the Rein-Sehgal Model \cite{reinsehgal}.
%\textbf{Shouldn't we say something about $\eta$ production in NEUT?}
More detailed information on the neutrino interaction and detector model are presented in~\cite{Abe2018}.

%% The atmospheric neutrino interactions with the nucleons in the SK tank are simulated by .
%%The neutrino interaction model has been updated since the latest paper \cite{Abe2017}, eventuating in more precise %background event rate expectation. The summary of the update can be found in \cite{Abe2018}.

%% This estimation is based on the measured primary cosmic ray fluxes in AMS \cite{Alcaraz2000} and BESS \cite{Alcaraz2000a} experiments with the effects of solar wind and geomagnetic field being taken into account.
%%Interactions of primary cosmic ray with air nuclei are simulated based on JAM \cite{Niita2006} for energies of the primary cosmic rays below 32~GeV and DPMJET-III \cite{Roesler2001} for above 32~GeV.
%Finally, the flux of atmospheric neutrinos is acquired from the decays of pions, kaons, and the secondary particles from air nuclei.

%%The simulation consists of atmospheric neutrino flux calculation, neutrino oscillation, and a detailed model of
%% neutrino-nucleus interaction cross sections.
%%Atmospheric neutrino background sample of 500-year detector exposure are generated for each SK period.

%The neutral pions induced by these interactions, which is the dominant background source, can interact with nucleons in \Ox~nucleus in a similar way to the $\eta$ nuclear effect: pions can be scattered, absorbed, or charge exchanged (final state interactions, FSI).

\section{\label{sec:reco}Data Reduction and\\ Event Reconstruction}

In this study data from a total exposure of 0.373 Mton$\cdot$years and corresponding to 6050.3 live days is analysed.
This analysis uses only Fully Contained (FC) events, whose reconstructed vertices and all visible particles are containd within the ID fiducial volume.
%% to suppress the number of background, mostly from the cosmic-ray muons. 
The fiducial volume is defined as the region 2~m away from the top, bottom, and barrel walls of the ID and corresponds to a 22.5~kton mass.
These FC events are extracted during multiple reduction steps by algorithms identical to those applied in atmospheric neutrino studies and other nucleon decay searches~\cite{Abe2017,Takenaka2020,Abe2018}.
After reduction, events are required to have vertices within the fiducial volume, with  visible energies greater than $30~\mathrm{MeV}$, and with no cluster of OD PMTs with more than 16 hits. 

%no cluster of hits is observed in the OD.

Reconstruction processes are applied to all FC events to determine the event kinematic parameters, such as the number of Cherenkov rings, their momenta, and their particle type (PID).
The APfit reconstruction~\cite{Shiozawa1999, TakenakaThesis2020} is used in this analysis, as in the other atmospheric neutrino analyses and nucleon decay searches, and is applied to both data and MC events.
The reconstruction starts by determining the vertex location using the point in the ID that maximizes the peak of the residual timing distribution assuming all light arriving at the PMTs originated from that point.
%%by finding the peak of timing residual ((photon arrival time)-(time of fight)) distribution of ID PMTs.
Next, the direction and the edge of the most energetic Cherenkov ring are estimated from the angular distribution of the observed charge as a function of the opening angle.
A ring pattern recognition algorithm based on the Hough transformation \cite{davies2004machine} is then used to search for additional Cherenkov rings.
Each ring is classified as either a showering particle $(e^\pm,~\gamma)$ or a non-showering particle $(\mu^\pm,~\pi^\pm)$ according to its hit pattern and opening angle.
While the former creates a blurred ring pattern due to the induced electromagnetic shower and multiple scattering of particles therein, the latter shows a sharper ring edge.
The momentum of each ring is evaluated using the observed charge inside the $70^{\circ}$ half-angle cone around the ring direction after accounting for light attenuation in water and the angular acceptance of the PMTs.
Michel electrons are tagged by searching for hit clusters
after the primary Cherenkov ring event.
%%The algorithm of event reconstruction is described in more detail in \cite{Shiozawa1999}.

%Thanks to the improvements of the 2008 electronics and data acquisition upgrade, the Michel electron tagging efficiency is higher and hence signal efficiency is improved in SK-IV, especially for nucleon decay modes including μþ in the final state. 

When neutrons thermalize in the detector water and capture on a hydrogen nucleus,  a 2.2~MeV $\gamma$-ray is emitted.
Such $\gamma$-rays can be tagged in the SK-IV period (and later) with a dedicated algorithm described in~\cite{Abe2022}.
The neutron tagging efficiency is estimated to be 25.2$\%$ with a false-positive rate of 1.8$\%$ based on atmospheric neutrino MC and AmBe calibration data. 
Since proton decay is expected to produce neutrons with a probability of less than 10$\%$ \cite{Tanaka2020, Ejiri1993}, neutron tagging enables the removal of many atmospheric neutrino backgrounds and hence improves the sensitivity of this search.

%% by searching seeking the hit clusters with more than five hit PMTs within a $10~\textrm{ns}$ sliding window after the prompt neutrino interaction for analyses of SK-IV period.
%16 variables that represent the characteristics of the candidate cluster are calculated and input into a neural network to classify into the real and fake clusters.
%% A set of 16 variables representing the characteristics of the candidate cluster is calculated and then input into a neural network to classify real and fake clusters.

In the event selection used below, the reconstructed total momentum, $P_\mathrm{tot}$, the invariant mass, $M_\mathrm{tot}$, and the total energy $E_\mathrm{tot}$ are defined as follows:
\begin{eqnarray}%
%\begin{align}
%\displaystyle    
    P_\mathrm{tot} &=& \left\vert\sum_{i=1}^{n_\mathrm{par}}\bm{p_i}\right\vert\label{eq:ptot}~,\\
    E_\mathrm{tot} &=& \sum_{i=1}^{n_\mathrm{par}}\sqrt{m_i^2+\bm{p_i}^2}\label{eq:etot}~,\\
    M_\mathrm{tot} &=& \sqrt{E_\mathrm{tot}^2 -P_\mathrm{tot}^2}\label{eq:mtot}~.
%    \end{align}
\end{eqnarray}
\noindent Here $m_i$ is the mass of the $i$-th particle of proton decay product, $\bm{p_i}$ is the momentum of $i$-th particle (ring), and $n_\mathrm{par}$ is the number reconstructed particles.

The reconstructed invariant mass of the $\eta$ meson, $m_\eta$, is similarly calculated by summing all the momenta and energies of the reconstructed particles identified as the products of the $\eta$ decay. 
This identification is made by taking all the possible combinations of observed showering Cherenkov rings and selecting the combination whose invariant mass reconstructs closest to the rest mass of $\eta$ meson.

\section{\label{sec:evsel}Event Selection}

The $\eta$ meson has three dominant decay modes with a lifetime of $5\times 10^{-19}~\textrm{s}$: $\eta\rightarrow 2\gamma$ (branching ratio of 39$\%$), $\eta\rightarrow 3\pi^0$ (branching ratio of 33$\%$), and $\eta\rightarrow\pi^+\pi^-\pi^0$ (branching ratio of 23$\%$).
The former two modes are analysed in this study, whereas the last one is excluded due to its smaller branching ratio and poor detection efficiency in SK \cite{Nishino2012}.

\subsection{$\eta\rightarrow 2\gamma$ search}
The event selection criteria for the search for $p\rightarrow e^+\eta, ~\eta\rightarrow 2\gamma$ and $p\rightarrow \mu^+\eta, ~\eta\rightarrow 2\gamma$ are defined as follows:
\begin{enumerate}
    \item[\textbf{(A1)}] There are three Cherenkov rings.
    \item[\textbf{(A2)}] All rings are showering for $p\rightarrow e^+\eta$ and one of the rings is non-showering for $p\rightarrow \mu^+\eta$.
    \item[\textbf{(A3)}] The reconstructed $\eta$ mass satisfies $480< M_\eta < 620~\textrm{MeV/}c^2$. 
    \item[\textbf{(A4)}] There is no tagged Michel electron for $p\rightarrow e^+\eta$ and one for $p\rightarrow \mu^+\eta$.
    \item[\textbf{(A5)}] The total momentum is $P_\mathrm{tot}<250~\textrm{MeV/}c$ and the invariant mass satisfies $800< M_\mathrm{tot} < 1050~\textrm{MeV/}c^2$.
    \item[\textbf{(A6)}] The total momentum $P_\mathrm{tot}$ is greater than or \textbf{(A7)} less than 100 MeV/$c$.
    \item[\textbf{(A8)}] There is no tagged neutron in the upper nor \textbf{(A9)} lower total momentum region for the SK-IV data.
\end{enumerate}

\begin{figure*}[]
\includegraphics[width=\linewidth]{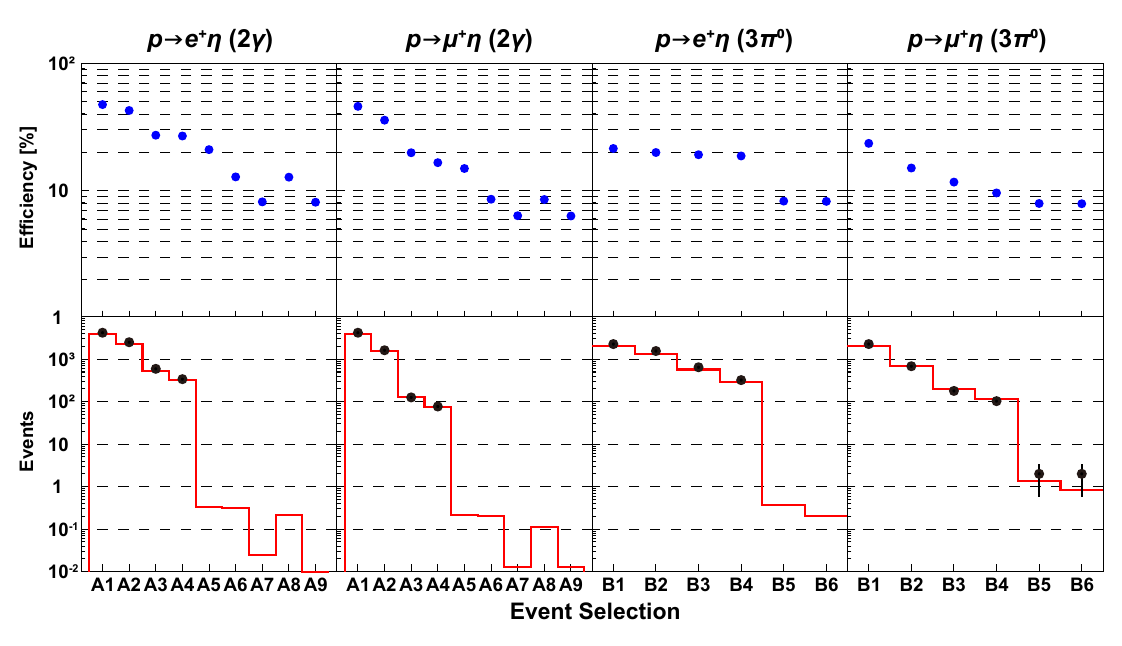}% Here is how to import EPS art
\caption{\label{fig:select_cut} The signal efficiencies (upper) and the number of expected backgrounds (lower, red histogram) and data candidates (lower, black circles) for $p\rightarrow e^+\eta~(\eta\rightarrow 2\gamma)$, $p\rightarrow \mu^+\eta~(\eta\rightarrow 2\gamma)$, $p\rightarrow e^+\eta~(\eta\rightarrow 3\pi^0)$, $p\rightarrow \mu^+\eta~(\eta\rightarrow 3\pi^0)$ searches from left to right. All the results from SK-I to SK-IV are combined. The event selection criteria are defined in Section~\ref{sec:evsel}. The numbers of events of backgrounds and data candidates agree within 15$\%$ from cuts A1-A4 and B1-B4, which are well covered by systematic model variations.}
\end{figure*}

%Fully Contained fiducial volume cut, which requires the distance between the reconstructed vertex and the wall to be larger than $200~\mathrm{cm}$, has been applied to all the events in the proton decay search. 
%The total momentum separation into two search boxes $(\textbf{A6}~\&~\textbf{A7})$ 
Two signal boxes are defined by selections on the total momentum $(\textbf{A6}$ \& $\textbf{A7})$: a lower signal box with $P_\mathrm{tot}<100~\textrm{MeV/}c$ and an upper signal box with $100 \leq P_\mathrm{tot}<250~\textrm{MeV/}c$ cuts.
This separation is applied since the expected background rate in the $p_\mathrm{tot} < 100~\mathrm{MeV/}c$ region is negligibly small ($\ll$ 0.1 events in each SK period) and signal events in this region are primarily from the decay of free protons.
Signal selection efficiencies and the expected number of atmospheric neutrino background events at each stage of the selection are shown in Fig.~\ref{fig:select_cut} and are summarized in Table~\ref{tab:sum_eff}.

%% Accordingly, events observed in the lower box are have a higher likelihood of as proton decay candidates.

\begin{table*}
\caption{\label{tab:sum_eff} Summary of the signal efficiencies, the number of expected background events and the number of candidate events from 91.5, 49.1, 31.8, and 199.9 kton$\cdot$years exposures of the SK-I, SK-II, SK-III, and SK-IV periods, respectively. The errors on the background events are statistical uncertainties from 500 years of atmospheric neutrino MC for each SK period. The ``upper" and ``lower" in the $\eta\rightarrow 2\gamma$ searches stand for $100\leq p_\mathrm{tot}<250~\textrm{MeV/}c$ and $p_\mathrm{tot}<100~\textrm{MeV/}c$, respectively.}
\begin{ruledtabular}
\begin{tabular}{ccccccccccccc}
&\multicolumn{4}{c}{Efficiency~[$\%$]}& \multicolumn{4}{c}{Background~[events]}& \multicolumn{4}{c}{Candidate [events]}\\\cline{2-5}\cline{6-9}\cline{10-13}
SK period & I & II & III & IV & I & II & III & IV & I & II & III &IV\\\hline
$p\rightarrow e^+\eta$\\
($2\gamma$, upper)& 13.9 & 11.9 & 12.9 & 12.2 & 0.08 $\pm$ 0.03 & 0.05 $\pm$ 0.01 &0.03 $\pm$ 0.01 & 0.05 $\pm$ 0.03 & 0 & 0 & 0 & 0\\
($2\gamma$, lower)& 8.2 & 7.8 & 8.5 & 8.0 & 0.000 $\pm$ 0.006 & 0.004 $\pm$ 0.004 &0.003 $\pm$ 0.003 & 0.000 $\pm$ 0.010 & 0 & 0 & 0 & 0\\
($3\pi^0$)& 8.4 & 8.4 & 7.7 & 8.4 & 0.06 $\pm$ 0.02 & 0.06 $\pm$ 0.02 &0.03 $\pm$ 0.01 & 0.05 $\pm$ 0.03 & 0 & 0 & 0 & 0\\\hline
$p\rightarrow \mu^+\eta$\\
($2\gamma$, upper)&8.3 & 6.9 & 8.9 & 9.9 & 0.05 $\pm$ 0.02 & 0.04 $\pm$ 0.01 &0.003 $\pm$ 0.003 & 0.03 $\pm$ 0.02 & 0 & 0 & 0 & 0\\
($2\gamma$, lower)&6.3 & 5.6 & 6.3 & 7.1 & 0.008 $\pm$ 0.008 & 0.004 $\pm$ 0.004 &0.000 $\pm$ 0.003 & 0.000 $\pm$ 0.010 & 0 & 0 & 0 & 0\\
($3\pi^0$)&7.9 & 6.2 & 7.8 & 9.6 & 0.38 $\pm$ 0.05 & 0.17 $\pm$ 0.03 &0.12 $\pm$ 0.02 & 0.17 $\pm$ 0.05 & 0 & 1 & 0 & 1
\end{tabular}
\end{ruledtabular}
\end{table*}

Efficiencies are evaluated using the number of proton decay MC events passing the selection criteria.
 Background events are estimated by applying the criteria to 500-year-equivalent atmospheric neutrino MC samples for each SK period (2000 years in total) and then normalizing by live time observed in data and 
reweighted based on the SK oscillation fit~\cite{Abe2018}.
% corrections applied based on fitted oscillation, flux, and cross section parameters from~\cite{Abe2018}.
In this analysis, signal selection efficiencies have increased by $\sim10\%$ compared to the previous analysis \cite{Abe2017} due to the re-estimation of $\eta$ nuclear effect as described in Section~\ref{sec:sim_nuc}.
The downward adjustment in $\sigma_\mathrm{nuc}$ around $p_\eta= 200-300~\textrm{MeV/}c$ as shown in Fig.~\ref{fig:xs_nuc} results in an increase in the number of proton decay signal events whose $\eta$ mesons have not been scattered nor absorbed, making them easier to detect.
%% as candidates.

The reconstructed momenta and mass cuts are especially significant in distinguishing proton decay signals from backgrounds as atmospheric neutrino events seldom produce isotropic event topologies; most neutrino-induced particles are emitted in the direction of the neutrino momentum. 
Selection ($\textbf{A5}$) reduces the number of expected background events to $<0.01$ integrated over all SK periods for both the $p\rightarrow e^+\eta$ and $p\rightarrow \mu^+\eta$ searches in the lower signal box analysis ($\textbf{A9}$), while keeping the signal efficiency high.
% above 1$\%$.

\begin{figure*}[]
\includegraphics[width=\linewidth]{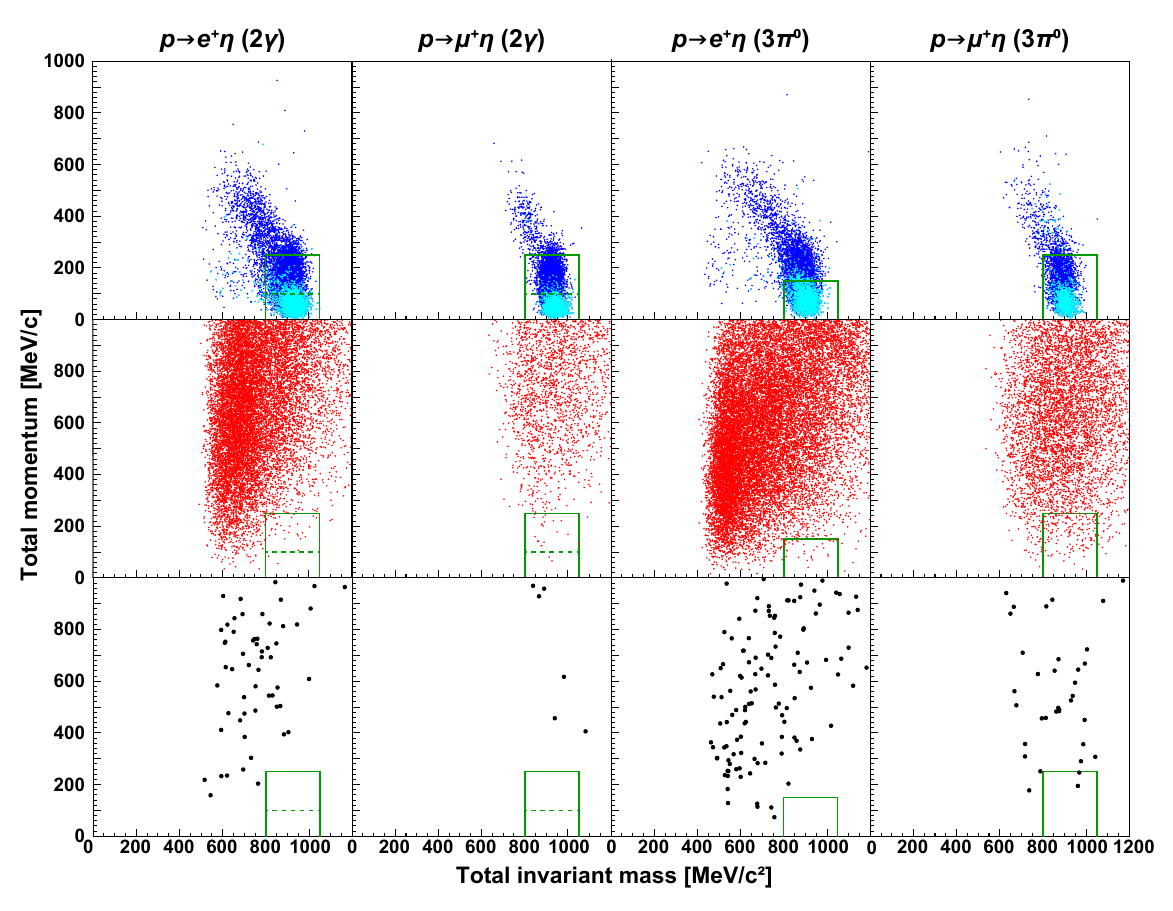}% Here is how to import EPS art
\caption{\label{fig:totmom_mass} Reconstructed total invariant mass and momentum of proton decay MC (upper), atmospheric neutrino MC (middle), and data (lower) for $p\rightarrow e^+\eta~(\eta\rightarrow 2\gamma)$, $p\rightarrow \mu^+\eta~(\eta\rightarrow 2\gamma)$, $p\rightarrow e^+\eta~(\eta\rightarrow 3\pi^0)$, $p\rightarrow \mu^+\eta~(\eta\rightarrow 3\pi^0)$ searches from left to right. The blue and cyan circles in the upper figures stand for bound and free proton decay events, respectively. The red circles in the middle and the black circles in the lower each correspond to the 500 years of atmospheric neutrino MC and the data from SK-I to SK-IV combined. The green solid and dashed lines indicate the selection criteria ($\textbf{A5-7}$) and ($\textbf{B5}$) in Section~\ref{sec:evsel}. All the event selections except ($\textbf{A5-A7}$) and ($\textbf{B5}$) are applied.}
\end{figure*}

\begin{figure*}[]
\includegraphics[width=\linewidth]{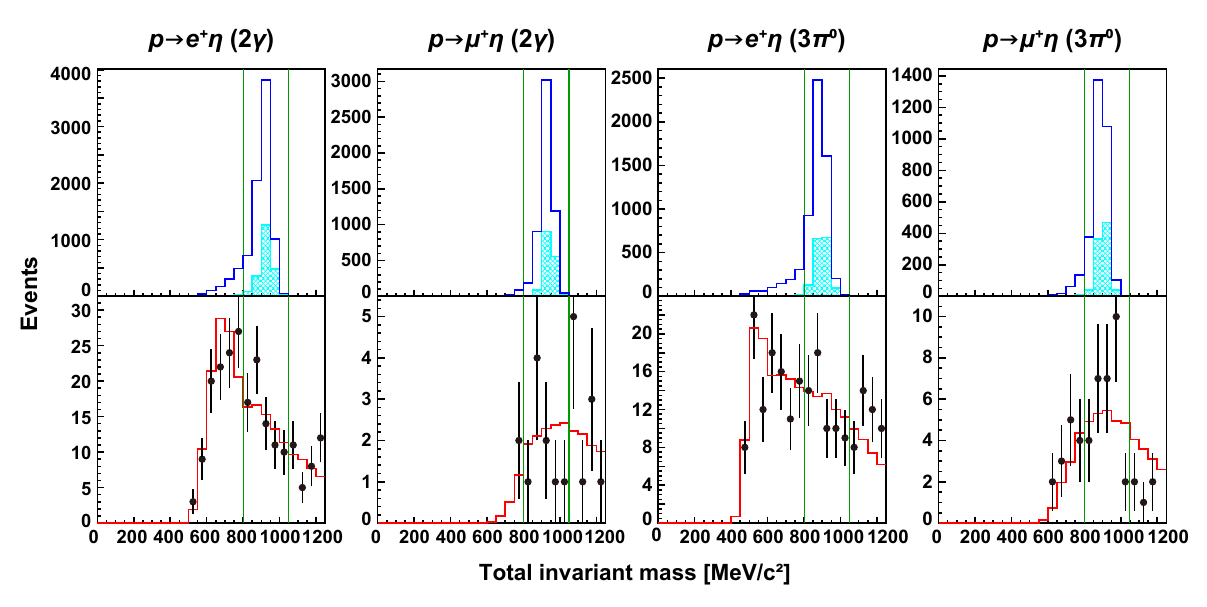}% Here is how to import EPS art
\caption{\label{fig:mass_dis} Reconstructed total invariant mass of proton decay MC (upper), atmospheric neutrino MC and data (lower) for $p\rightarrow e^+\eta~(\eta\rightarrow 2\gamma)$, $p\rightarrow \mu^+\eta~(\eta\rightarrow 2\gamma)$, $p\rightarrow e^+\eta~(\eta\rightarrow 3\pi^0)$, $p\rightarrow \mu^+\eta~(\eta\rightarrow 3\pi^0)$ searches from left to right. The blue and cyan histograms in the upper figures stand for bound and free proton decay events, respectively. The red lines and the black circles in the lower each correspond to the 500 years of atmospheric neutrino MC and the data from SK-I to SK-IV combined. The green solid and dashed lines indicate the selection criteria ($\textbf{A5-7}$) and ($\textbf{B5}$) in Section~\ref{sec:evsel}. All the event selections except ($\textbf{A5-A7}$) and ($\textbf{B5}$) are applied.}
\end{figure*}

\begin{figure*}[]
\includegraphics[width=\linewidth]{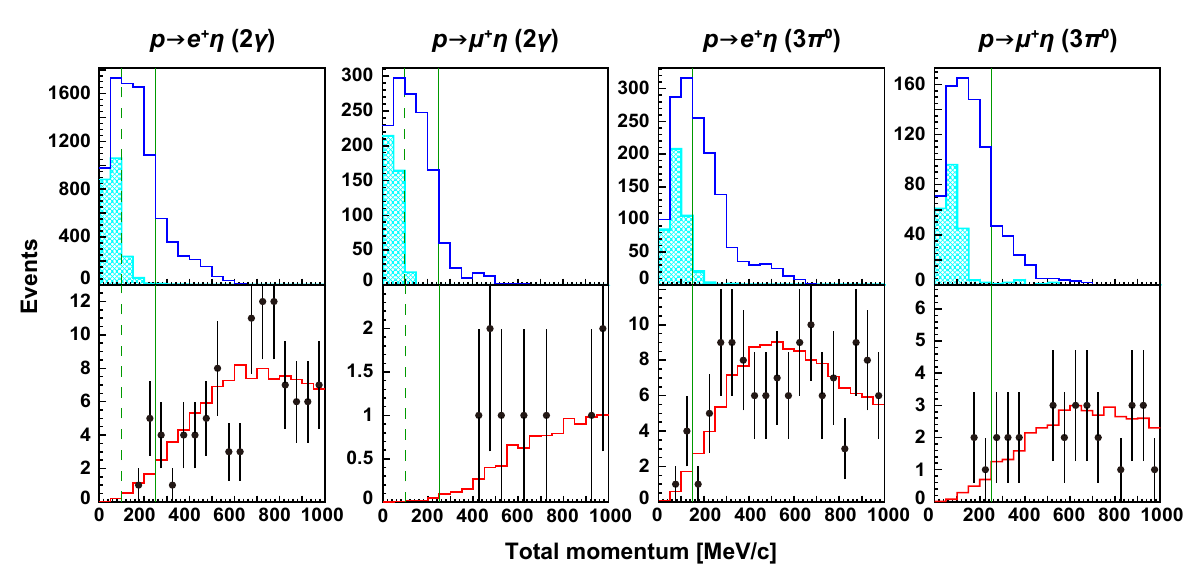}% Here is how to import EPS art
\caption{\label{fig:tmom_dis} Reconstructed total momentum of proton decay MC (upper), atmospheric neutrino MC and data (lower) for $p\rightarrow e^+\eta~(\eta\rightarrow 2\gamma)$, $p\rightarrow \mu^+\eta~(\eta\rightarrow 2\gamma)$, $p\rightarrow e^+\eta~(\eta\rightarrow 3\pi^0)$, $p\rightarrow \mu^+\eta~(\eta\rightarrow 3\pi^0)$ searches from left to right. The blue and cyan histograms in the upper figures stand for bound and free proton decay events, respectively. The red lines and the black circles in the lower each correspond to the 500 years of atmospheric neutrino MC and the data from SK-I to SK-IV combined. The green solid and dashed lines indicate the selection criteria ($\textbf{A5-7}$) and ($\textbf{B5}$) in Section~\ref{sec:evsel}. All the event selections except ($\textbf{A5-A7}$) and ($\textbf{B5}$) are applied.}
\end{figure*}

Figure~\ref{fig:totmom_mass} shows the distribution of events passing the selection in the total mass and total momentum plane, but excluding cuts on those variables. 
One-dimensional distributions of these parameters are shown in Fig.~\ref{fig:mass_dis} and Fig.~\ref{fig:tmom_dis} for the signal MC, the 2000-year atmospheric neutrino MC, and the integrated data from SK-I to SK-IV.
All selection criteria except for those on the plotted variables are applied.
Most of the free proton decay events populate the lower signal boxes ($P_\textrm{tot}<100~\textrm{MeV/}c$).

\subsection{$\eta\rightarrow 3\pi^0$ search}
The event selection criteria for the search for $p\rightarrow e^+\eta,~\eta\rightarrow 3\pi^0$ and $p\rightarrow \mu^+\eta,~\eta\rightarrow 3\pi^0$ are defined as follows:
\begin{enumerate}
    \item[\textbf{(B1)}] There are four or five Cherenkov rings.
    \item[\textbf{(B2)}] All rings are showering for $p\rightarrow e^+\eta$ and one of the rings is non-showering for $p\rightarrow \mu^+\eta$.
    \item[\textbf{(B3)}] The reconstructed $\eta$ mass satisfies $400< M_\eta < 700~\textrm{MeV/}c^2$. 
    \item[\textbf{(B4)}] There is no tagged Michel electron for $p\rightarrow e^+\eta$ and one for $p\rightarrow \mu^+\eta$.
    \item[\textbf{(B5)}] The total momentum is $P_\mathrm{tot}<150~\textrm{MeV/}c$ for $p\rightarrow e^+\eta$ and $P_\mathrm{tot}<250~\textrm{MeV/}c$ for $p\rightarrow \mu^+\eta$. The invariant mass satisfies $800< M_\mathrm{tot} < 1050~\textrm{MeV/}c^2$.
    \item[\textbf{(B6)}] There is no tagged neutron for the SK-IV data.
\end{enumerate}

The signal selection efficiencies and the expected number of atmospheric neutrino background events at each step of the selection are shown in Fig.~\ref{fig:select_cut}.
In this decay mode the three neutral pions decay into six $\gamma$-rays and generate six Cherenkov rings. 
However, as the ring counting algorithm is only able to identify up to five rings, the condition on the number of rings is set to be four or five ($\mathbf{B1}$).
This results in a reduced ability to reconstruct the $\eta$ mass and therefore the event selection window of $\eta$ mass $(\textbf{B3})$ is set to be larger than in the $\eta\rightarrow 2\gamma$ search.
The two-dimensional total mass and total momentum distributions and the one-dimensional distributions of the total mass and total momentum are shown in Figs.~\ref{fig:totmom_mass},~\ref{fig:mass_dis}, and ~\ref{fig:tmom_dis}, respectively. 
Unlike the $\eta\rightarrow 2\gamma$ search, a two-box separation in total momentum is not implemented here as a substantial fraction of events derived from free proton decays are reconstructed with total momenta greater than 100 MeV$/c$ as shown in the upper figures in Fig.~\ref{fig:tmom_dis}.
A tighter cut on the total momentum of $P_\textrm{tot}<150~\textrm{MeV}/c$ at $(\textbf{B5})$ is applied for $p\rightarrow e^+\eta$ mode to reduce backgrounds.
The breakdown of the remaining background events by neutrino interaction mode is listed in Table~\ref{tab:bg_brkdwn}.
Atmospheric neutrino events with neutral pions generating multiple showering rings dominate the background in both modes.

\begin{table}
\caption{\label{tab:bg_brkdwn} Breakdown (percentage contribution) of the neutrino interaction modes of the background events in the final signal boxes from SK-I to SK-IV in total. Here, CC and NC represent charged-current and neutral-current, respectively. QE, 1$\pi$, and multi$\pi$
stand for quasielastic scattering, single $\pi$ production and multiple $\pi$ production, respectively.}
\begin{ruledtabular}
\begin{tabular}{ccc}
Interaction&$p\rightarrow e^+\eta$&$p\rightarrow \mu^+\eta$\\\hline
%&$\eta\rightarrow 2\gamma$&$\eta\rightarrow 2\gamma$\\
CCQE & 6 &1 \\
CC1$\pi$ & 19&14\\
CC multi$-\pi$& 25 & 36\\
CC~$\&$~NC 1$\eta$ & 6 & 11\\
CC others & 7 &8 \\
NC& 37 &20
\end{tabular}
\end{ruledtabular}
\end{table}

\section{\label{sec:result}Search Results}
The search for proton decay was performed using $0.37~\mathrm{Mton\cdot years}$ of SK data by applying the selection criteria described in Section~\ref{sec:evsel}.
No signal candidate was found for the $p\rightarrow e^+\eta$ mode whereas two events remain in the final signal region of the $\eta\rightarrow 3\pi^0$ decay search in the $p\rightarrow\mu^+\eta$ mode.
Both of the candidates are near the selection thresholds of the total mass and momentum cuts and 
are identical to those found in the previous analysis \cite{Abe2017}.
No significant data excess was found above the expected number of atmospheric neutrino background events in either mode.
The total estimated background is $0.42$ and $0.93$ events for the $p\rightarrow e^+\eta$ and $p\rightarrow \mu^+\eta$ modes, respectively.
The Poisson probability to observe zero (two) or more events assuming mean values of 0.42 (0.93) is 65.7$\%$ (23.9$\%$).
As shown in Fig.~\ref{fig:select_cut}, Fig.~\ref{fig:mass_dis}, and Fig.~\ref{fig:tmom_dis}, the data and background MC are consistent both in the event rates along the event selection cuts and in the distributions of the relevant selection parameters.

\subsection{\label{sec:result_syserr}Systematic Uncertainties}

Systematic errors on the signal efficiencies and on the number of expected background events are summarized in Tables~\ref{tab:sum_sys_sig} and \ref{tab:sum_sys_bg}.
Uncertainties from the physics modelling and event reconstruction performance are considered for both the proton decay signal and the atmospheric neutrino background MC.
These systematic errors are estimated using the same methods as in the previous paper \cite{Abe2017} except for those coming from $\eta$ and $\pi$ nuclear effects, as well as those from $\pi$ secondary interactions in water for the background.

\begin{table*}
\caption{\label{tab:sum_sys_sig} Summary of the systematic uncertainties on signal efficiencies for each factor (in units of percent). 
The errors are averaged over SK-I
to SK-IV periods by live time. The "upper" and "lower" in the $\eta\rightarrow 2\gamma$ searches stand for $100\leq p_\mathrm{tot}<250~\textrm{MeV/}c$ and $p_\mathrm{tot}<100~\textrm{MeV/}c$ regions, respectively.}
\begin{ruledtabular}
\begin{tabular}{cccccc}
& $\eta$ nuclear & N-N correlated & Fermi & Event\\
Modes & effect & decay & momentum & reconstruction & Total\\\hline
$p\rightarrow e^+\eta$\\
($2\gamma$, upper)& 9.1 & 9.6 & 6.0 & 4.6 & 14.7  \\
($2\gamma$, lower)& 7.0 & 3.1 & 13.2 & 4.3 & 15.5 \\
($3\pi^0$)& 6.6 & 5.1 & 15.5 & 3.7 & 17.9 \\\hline
$p\rightarrow \mu^+\eta$\\
($2\gamma$, upper)& 8.4 & 9.8 & 7.5 & 3.6 & 15.6 \\
($2\gamma$, lower)& 8.2 & 3.0 & 13.9 & 3.2 & 16.8 \\
($3\pi^0$)& 9.4 & 6.8 & 2.9 & 6.1 & 13.8
\end{tabular}
\end{ruledtabular}
\end{table*}

\begin{table*}
\caption{\label{tab:sum_sys_bg} Summary of the systematic uncertainties on the number of background events for each factor (in units of percent). The errors are averaged over SK-I
to SK-IV periods by live time. The "upper" and "lower" in the $\eta\rightarrow 2\gamma$ searches stand for $100\leq p_\mathrm{tot}<250~\textrm{MeV/}c$ and $p_\mathrm{tot}<100~\textrm{MeV/}c$ regions, respectively.}
\begin{ruledtabular}
\begin{tabular}{ccccccc}
& Neutrino & Neutrino &  $\pi$ & $\eta$ & Event&\\
Modes & flux & cross section  & interaction & nuclear effect & reconstruction & Total\\\hline
$p\rightarrow e^+\eta$\\

($2\gamma$, upper)& \multirow{2}{*}{8.5} & \multirow{2}{*}{15.6} & \multirow{2}{*}{8.9} & \multirow{2}{*}{3.0} & \multirow{2}{*}{24.1} & \multirow{2}{*}{31.6} \\
($2\gamma$, lower)&  &  &  &  &  & \\
($3\pi^0$)& 8.8 & 14.9 & 13.2 & 1.0 & 18.8 & 28.9   \\\hline
$p\rightarrow \mu^+\eta$\\
($2\gamma$, upper)& \multirow{2}{*}{9.3} & \multirow{2}{*}{20.6} & \multirow{2}{*}{14.9} & \multirow{2}{*}{3.0} & \multirow{2}{*}{29.1} & \multirow{2}{*}{39.9}  \\
($2\gamma$, lower)&  &  &  &  &  &  \\
($3\pi^0$)& 8.8 & 10.4 & 15.7 & 1.0 & 10.2 & 23.0
\end{tabular}
\end{ruledtabular}
\end{table*}

%These systematic errors are estimated in the same methods as in the previous paper \cite{Abe2017} except for those coming from $\eta$ nuclear effect and $\pi$ FSI and secondary interactions (SI) in water of background.

The systematic uncertainty arising from the $\eta$ nuclear effect was the dominant error source for signal efficiency in the previous analysis \cite{Abe2017}.
%In this study, this uncertainty is estimated by varying the fitting parameters of its cross section $\sigma_\textrm{nuc}$ within $\pm1\sigma$ uncertainties of experimental data of $\eta$ photoproduction reaction and $\eta$ absorption reaction from \cite{Robig-Landau1996} (see Section~\ref{sec:sim_nuc}).
In this study the uncertainty is estimated by varying the fitting parameters used to model $\sigma_\textrm{nuc}$ within their $\pm1\sigma$ errors (see Section~\ref{sec:sim_nuc}) and propagating the altered cross section through the full analysis.
%% obtained by the contour of $\chi^2$ value of simulation and experimental data of $\eta$ photoproduction reaction from \cite{Robig-Landau1996} .
The variation in the final signal efficiency relative to that from the nominal MC is adopted as the uncertainty.
%This new method provides a more reliable evaluation of systematic uncertainty from $\eta$ nuclear interactions compared to the previous analysis, leading to a reduction in the associated error.
This new method provides a more reliable evaluation of systematic uncertainty from $\eta$ nuclear interactions leading to a reduction in the associated error compared to the previous analysis, where the uncertainty in $\sigma_\mathrm{nuc}$ was evaluated
by scaling the Breit Wigner function to cover all the measured data points in \cite{Robig-Landau1996}.
For the upper momentum box of the $\eta\rightarrow 2\gamma$ search, where the majority of the signal candidates derive from bound protons, this is a factor of three reduction.
The error on the number of background events is taken as the variation in the event rate between MC generated with and without the updated nuclear effect cross section.

%In the previous study, $\pi$ FSI and SI uncertainties were considered to be independent and estimated respectively.
In the previous study, the $\pi$ nuclear effect and secondary interaction uncertainties were considered to be independent and estimated separately.
However, since these two uncertainties have common error sources, they are now evaluated as a single error in this study. 
The current approach is identical to that adopted in the latest $p\rightarrow e^+\pi^0$ search at SK \cite{Takenaka2020}.
This improvement results in a reduction of the systematic error by at most $\sim$30$\%$ compared to the previous $p\rightarrow l^+\eta$ analysis.
%The errors arising from disagreement between data and MC simulations are assessed by the largest deviations between the numbers of atmospheric neutrino backgrounds and data through cuts A1-A4 and B1-B4.

%relatively smaller systematic uncertainties than the previous analysis.

There are seven sources of uncertainty stemming from the event reconstruction: uncentainty in the vertex position, the number of Cherenkov rings, the particle identification (ring pattern), the number of Michel electrons, the energy scale, and the number of tagged neutrons for the SK-IV period.
The same error estimation methods are applied for both the signal selection efficiencies and the number of expected background events.
Due to limited MC statistics, the upper and lower signal boxes for the atmospheric neutrino background are combined when estimating the uncertainty. 
The resulting background error is then applied to both signal boxes.

\subsection{\label{sec:result_lifetime}Lifetime Limits}
No significant event excess was observed in either decay mode and lower limits on the proton's partial lifetime are calculated using a Bayesian method \cite{AMSLER20081, PhysRevD.63.013009}.
Lifetime limits are calculated by combining the 12 independent measurements for each $p\rightarrow l^+\eta$ mode: two (upper and lower) box analyses for $\eta\rightarrow 2\gamma$ and single signal box analysis for $\eta\rightarrow 3\pi^0$ taken over four different SK periods each.

The probability density function for each proton decay rate, $\Gamma$, is defined as

\begin{eqnarray}
    \mathbf{P}(\Gamma|n)=&&\frac{1}{A_i}\iiint \frac{e^{-(\Gamma\lambda_i\epsilon_i+b_i)}(\Gamma\lambda_i\epsilon_i+b_i)^n_i}{n_i!}\nonumber\\
    &&\times\mathbf{P}(\Gamma)\mathbf{P}(\lambda_i)\mathbf{P}(\epsilon_i)\mathbf{P}(b_i)d\epsilon_i d\lambda_i db_i~,
\end{eqnarray}

\noindent where $i$ is the index of each measurement, $n_i$ is the number of data candidates, $\lambda_i$ is the detector exposure, $\epsilon$ is the signal selection efficiency, and $b_i$ is the expected number of background events.
The prior probability distribution on the decay rate $\mathbf{P}(\Gamma)$ is $1$ for $\Gamma\geq 0$ and otherwise $0$. $\mathbf{P}(\lambda_i)$, $\mathbf{P}(\epsilon_i)$, and $\mathbf{P}(b_i)$ stand for the prior probabilities for the detector exposure, signal efficiency, and number of background events, respectively, which are assumed to be Gaussian distributions:

\begin{eqnarray}
    \mathbf{P}(\lambda_i)\propto \left\{ \begin{array}{ll}
    \exp{\left(-\dfrac{(\lambda_i-\lambda_{0i})^2}{2\sigma_{\lambda_i}^2}\right)}, & (\lambda_i > 0)\\
    0, & (\lambda_i\leq 0)
  \end{array} \right.
\end{eqnarray}

\begin{eqnarray}
    \mathbf{P}(\epsilon_i)\propto \left\{ \begin{array}{ll}
    \exp{\left(-\dfrac{(\epsilon_i-\epsilon_{0i})^2}{2\sigma_{\epsilon_i}^2}\right)}, & (\epsilon_i > 0)\\
    0, & (\epsilon_i\leq 0)
  \end{array} \right.
\end{eqnarray}

\noindent Here $\lambda_{0i}~(\sigma_{\lambda_i})$ and $\epsilon_{0i}~(\sigma_{\epsilon_i})$ are nuissance parameters for the exposure and signal efficiency (their uncertainties).
The uncertainties on the detector exposures are assumed to be less than 1$\%$. 
As the number of background events $b_i$ is low enough, a convolution of Poisson and Gaussian distribution is adopted for the probability of the number of background to take into account the MC statistical error.
Therefore, $\mathbf{P}(b)$ is expressed as
\begin{eqnarray}
    \mathbf{P}(b_i)\propto \left\{\begin{array}{ll}
    \begin{aligned}
    \int_0^\infty&\frac{e^{-B} (B)^n_{b_i}}{n_{b_i}!}\\
    \times&\exp{\left(-\frac{({b_i}{C_i}-B)^2}{2\sigma_{b_i}^2}\right)}dB,& (b_i > 0)
    \\
    0.&&(b_i\leq 0)
    \end{aligned} %& (b_i > 0)\\
    %0. & (b_i\leq 0)
  \end{array} \right.
\end{eqnarray}
Here $n_{b_i}$ is the expected number of background events in 500 years of atmospheric neutrino MC, $B$ is the number of true background events in 500 years of atmospheric neutrino MC, $C_i$ is the normalization constant for MC live time to the data live time, and $\sigma_{b_i}$ is the systematic error of the number of background events.
With these definitions, the lower limit on the proton decay rate for the full search, $\Gamma_\textrm{limit}$, at a given $\textrm{C.L.}$ is
\begin{eqnarray}
    \textrm{C.L.} =\int_0^{\Gamma_\mathrm{limit}}\prod_{i=1}^{12}\mathbf{P_i}(\Gamma|n_i)~d\Gamma,
\end{eqnarray}
\noindent with the lower lifetime limit defined as:
\begin{eqnarray}
    \tau/B = \frac{1}{\Gamma_\mathrm{limit}}.
\end{eqnarray}

The lifetime limits at $90\%$ $\textrm{C.L.}$ are $\tau/B(p\rightarrow e^+\eta) > 1.4\times 10^{34}~\mathrm{years}$ and $\tau/B(p\rightarrow \mu^+\eta) > 7.3\times 10^{33}~\mathrm{years}$.
Both limits have increased by a factor of 1.5 relative to the previous analysis.
This increase is attributed to both the newly analysed SK-IV data (55.7~kton$\cdot$years) which corresponds to a $\sim18\%$ increase in the total exposure and to the revised $\eta$ nuclear effect, which has improved the signal efficiency and reduced associated systematic errors. 
The results of these searches are summarized in Table~\ref{tab:lifetime}.

\begin{table}[]
\caption{\label{tab:lifetime} Summary of proton decay search of $p\rightarrow l^+\eta$.}
\begin{ruledtabular}
\begin{tabular}{ccccc}
& Background & Candidate & Probability & Lifetime limit \\
Modes & [events] & [events] & [$\%$] & at 90$\%$ CL\\\hline
$p\rightarrow e^+\eta$ & 0.42 $\pm$ 0.13 & 0 & 65.7& $1.4\times\mathrm{10^{34}~yrs.}$\\
$p\rightarrow \mu^+\eta$& 0.93 $\pm$ 0.25 & 2 & 23.9 & $7.3\times\mathrm{10^{33}~yrs.}$
\end{tabular}
\end{ruledtabular}
\end{table}

\section{\label{sec:concl}Conclusion}

Searches for proton decay into a charged antilepton and an $\eta$ meson were performed with an updated nuclear effect estimation and with an additional 55.7~kton$\cdot$years of SK-IV data since the last analysis \cite{Abe2017}.
The updated nuclear effect led to improvements both in signal efficiency ($\sim10$\%) and a reduction in this previously dominantly systematic uncertainty (a factor of three).
Using a combined exposure of $0.37~\mathrm{Mton\cdot years}$, the number of candidate events was consistent with the atmospheric neutrino background prediction and no indication of proton decay was observed for either mode.
Lower limits on the partial proton lifetime of $1.4\times\mathrm{10^{34}~years}$ for $p\rightarrow e^+\eta$ and $7.3\times\mathrm{10^{33}~years}$ for $p\rightarrow \mu^+\eta$ are set at 90$\%$ C.L.
These limits are around 1.5 times longer than our previous study and are the most stringent to date.

%%These results set the most stringent limits in the world.

\begin{acknowledgments}
%Natsumi Taniuchi has been supported by JSPS Grant-in-Aid for JSPS Research Fellows JP21J20623.
We gratefully acknowledge the cooperation of the Kamioka Mining and Smelting Company.
The Super-Kamiokande experiment has been built and operated from funding by the 
Japanese Ministry of Education, Culture, Sports, Science and Technology; the U.S.
Department of Energy; and the U.S. National Science Foundation. Some of us have been 
supported by funds from the National Research Foundation of Korea (NRF-2009-0083526,
NRF-2022R1A5A1030700, NRF-2202R1A3B1078756) funded by the Ministry of Science, 
Information and Communication Technology (ICT); the Institute for 
Basic Science (IBS-R016-Y2); and the Ministry of Education (2018R1D1A1B07049158,
2021R1I1A1A01042256, 2021R1I1A1A01059559); the Japan Society for the Promotion of Science; the National
Natural Science Foundation of China under Grants No.12375100; the Spanish Ministry of Science, 
Universities and Innovation (grant PID2021-124050NB-C31); the Natural Sciences and 
Engineering Research Council (NSERC) of Canada; the Scinet and Westgrid consortia of
Compute Canada; 
the National Science Centre (UMO-2018/30/E/ST2/00441 and UMO-2022/46/E/ST2/00336) 
and the Ministry of  Science and Higher Education (2023/WK/04), Poland;
the Science and Technology Facilities Council (STFC) and
Grid for Particle Physics (GridPP), UK; the European Union's 
Horizon 2020 Research and Innovation Programme under the Marie Sklodowska-Curie grant
agreement no.754496; H2020-MSCA-RISE-2018 JENNIFER2 grant agreement no.822070, H2020-MSCA-RISE-2019 SK2HK grant agreement no. 872549; 
and European Union's Next Generation EU/PRTR  grant CA3/RSUE2021-00559; 
the National Institute for Nuclear Physics (INFN), Italy.

%We gratefully acknowledge the cooperation of the Kamioka Mining and Smelting Company. The Super-Kamiokande experiment has been built and operated from funding by the Japanese Ministry of Education, Culture, Sports, Science and Technology, the U.S. Department of Energy, and the U.S. National Science Foundation. Some of us have been supported by funds from the National Research Foundation of Korea (NRF-2009-0083526 and NRF 2022R1A5A1030700) funded by the Ministry of Science, ICT, the Institute for Basic Science (IBS-R016-Y2), and the Ministry of Education (2018R1D1A1B07049158, 2021R1I1A1A01042256, 2021R1I1A1A01059559), the Japan Society for the Promotion of Science, the National Natural Science Foundation of China under Grants No. 11620101004, the Spanish Ministry of Science, Universities and Innovation (grant PGC2018-099388-BI00), the Natural Sciences and Engineering Research Council (NSERC) of Canada, the Scinet and Westgrid consortia of Compute Canada, the National Science Centre (UMO-2018/30/E/ST2/00441) and the Ministry of Education and Science (DIR/WK/2017/05), Poland, the Science and Technology Facilities Council (STFC) and GridPPP, UK, the European Union’s Horizon 2020 Research and Innovation Programme under the Marie Sklodowska-Curie grant agreement no. 754496, H2020-MSCA-RISE-2018 JENNIFER2 grant agreement no.822070, H2020-MSCARISE-2019 SK2HK grant agreement no. 872549.
\end{acknowledgments}

\bibliography{myreference}% Produces the bibliography via BibTeX.

\end{document}